\documentclass[aps,prl,twocolumn,groupedaddress]{revtex4-2}

\usepackage{graphicx}
\usepackage{xcolor}

\begin{document}



\title{When Is Structural Lubricity Load Independent? The Role of Contact Geometry and Elastic Compliance}


\author{Hongyu Gao}
\email[]{hongyu.gao@uni-saarland.de}
\affiliation{Department of Materials Science \& Engineering, Saarland University, Campus C6.3, 66123 Saarbrücken, Germany}


\date{\today}

\begin{abstract}

Using molecular dynamics simulations of an incommensurate Au(111)/graphite interface, we investigate the conditions under which structural lubricity produces load-independent friction.
We show that strict load independence occurs only in laterally infinite, area-filling contacts, where dissipation is governed by phonon-mediated viscous coupling and the shear stress scales linearly with sliding velocity. 
Finite contacts with explicit boundary terminations exhibit substantially higher friction yet remain load independent up to a critical load.
Load dependence arises only when elastic out-of-plane deformation near the contact line exceeds a critical amplitude, activating additional dissipation channels. 
These results demonstrate that contact geometry and local elastic compliance, rather than normal load itself, determine the onset and breakdown of load-independent structural lubricity.

\end{abstract}


\maketitle

Friction is traditionally described by the Amontons-Coulomb law~\cite{Bowden2001,Persson2000}, which states that the friction force scales linearly with the applied normal load. 
At macroscopic scales, this behavior emerges from the collective response of numerous microscopic asperity contacts. 
Increasing load engages additional junctions and strengthens existing ones, resulting in a proportional increase in real contact area and shear resistance. 
Although elastic contact mechanics predicts a sublinear growth of contact area with load~\cite{Hertz1882,Muser2001PRL,Dirk2013PRL}, and adhesion allows finite contact even at zero load, friction remains load dependent because normal load fundamentally modifies the contact geometry through changes in the number, size, or configuration of load-bearing asperities.

At the atomic scale, single-asperity friction is commonly described within the Prandtl~\cite{Prandtl1928ZAMM}(-Tomlinson)~\cite{Popov2012ZAMM} framework.
In this picture, a tip (apex) traverses a laterally corrugated potential energy surface at essentially fixed contact area.
Increasing the normal load enhances the corrugation amplitude and raises the associated energy barriers to slip, thereby increasing friction.
Macroscopic multicontact interfaces can be viewed as ensembles of such elementary asperities.
Although the underlying mechanisms differ--contact geometry evolution at large scale versus energy landscape modulation at the nanoscale--classical friction remains inherently load dependent across scales.

A distinct paradigm arises when these conventional load-controlled mechanisms are suppressed.
Structural lubricity~\cite{Mueser2004EL} occurs at clean, defect-free, atomically smooth crystalline interfaces in incommensurate contact, such as noble metals sliding on graphene.
In this regime, lateral force contributions from individual atoms cancel due to lattice mismatch, resulting in ultralow friction~\cite{Martin1993PRB,Dienwiebel2004PRL}.
Prior studies have reported dramatic friction (coefficient) reduction and sublinear scaling with contact size~\cite{Dirk2013PRL,Koren2016PRB}, persisting even in the presence of adsorbates~\cite{Cihan2016NC,Oo2024NL}. 
Yet whether such extensive force cancellation can produce genuine load-independent friction remains an open question. 
In realistic finite contacts, normal load may still induce dissipation through edge pinning~\cite{He1999S}, local registry variations~\cite{deWijn2012PRB}, or enhanced out-of-plane compliance~\cite{Sharp2016PRB}, thereby reintroducing load sensitivity even in nominally incommensurate systems.

Here we address this question by isolating contact geometry as an independent control variable within a structurally lubric interface.
We contrast a laterally infinite, area-filling contact, wherein boundary effects and contact-area evolution are eliminated by construction, with finite contacts that contain explicit termination lines.
This minimal framework allows the influence of normal load to be disentangled from geometric constraints.
It thereby enables a direct examination of how load sensitivity emerges or is suppressed in incommensurate crystalline contacts, without invoking disorder, plasticity, or chemical heterogeneity.

\begin{figure}[ht]
\centering    
\includegraphics[width=1.0\linewidth]{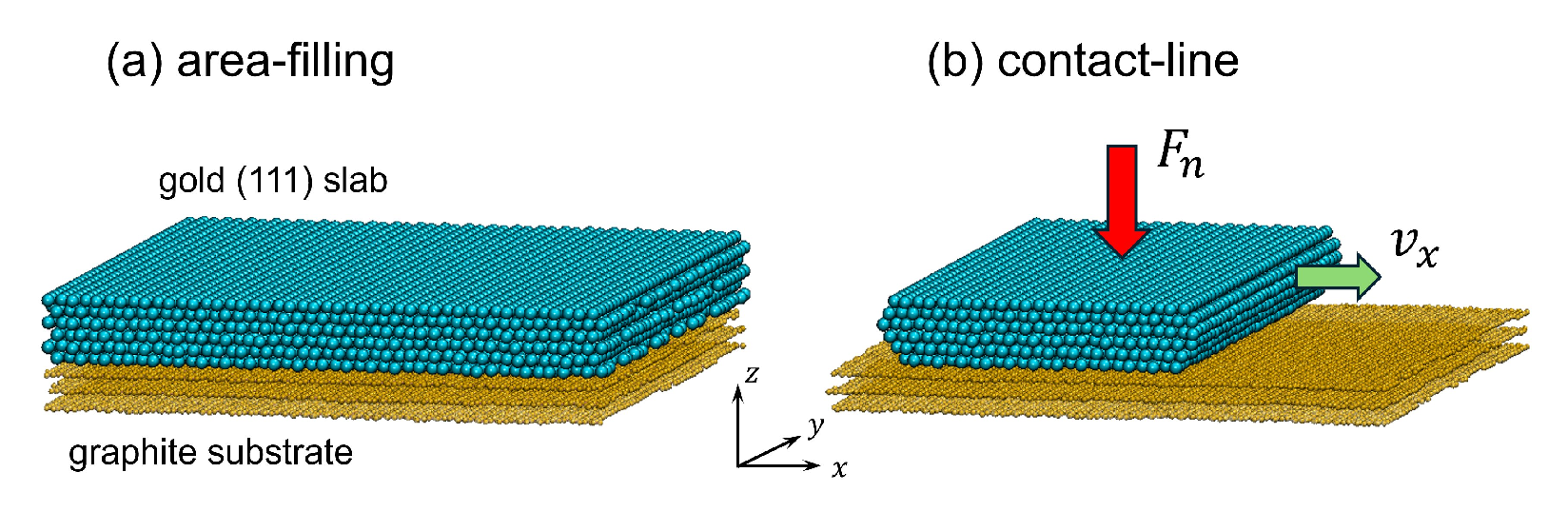}
\caption{
Simulation models of (a) the area-filling and (b) contact-line geometries.
The graphite substrate has in-plane dimensions of 11.5$\times$11.6 nm$^2$ with periodic boundary conditions applied laterally.
The topmost layer of the Au(111) slab is treated as rigid, with the normal load $F_n$ applied uniformly, and the slab is driven at a constant velocity $v_x$.
The remaining gold atoms and the top graphene layer evolve freely without positional constraints.
}
\label{fig:model}
\end{figure}

We perform molecular dynamics (MD) simulations of a gold slider in contact with a graphite substrate, as illustrated in Fig~\ref{fig:model}.
The substrate consists of three graphene layers, with the armchair direction aligned along the sliding direction.
The gold slab exposes a (111) surface parallel to the interface, forming an incommensurate contact with the underlying graphite.
Two contact geometries are considered:
(i) a laterally periodic, area-filling configuration representing an effectively infinite interface, and (ii) a finite geometry with termination lines oriented normal to the sliding direction.
In both cases, the gold slab is driven at a constant velocity $v_x$ (0.1--100 m/s) under a prescribed normal load $F_n$ spanning adhesive conditions up to 1 GPa.
The temperature is maintained at 300 K using a Langevin thermostat applied to the middle graphene layer with a damping time of 0.1 ps.

Interatomic interactions within the gold slab and graphite substrate are described using an embedded-atom method (EAM) potential~\cite{Zhou2004PRB} and the AIREBO potential~\cite{Stuart2000JCP}, respectively.
The gold-graphite interaction is modeled using a Morse potential~\cite{Rosa-Abad2016RSC}, parameterized to reproduce the equilibrium adhesion and interfacial separation.
Each simulation extends over a total sliding distance exceeding 200 nm, which is sufficient to ensure convergence to steady-state frictional behavior.
Friction forces are evaluated during steady sliding and averaged to obtain mean shear stresses, with statistical uncertainties estimated from temporal fluctuations.
All simulations are carried out using the open-source MD package LAMMPS~\cite{Thompson2022CPC}.

In the area-filling (AF) geometry, the laterally periodic Au(111)/\allowbreak graphite interface exhibits smooth sliding characteristic of structural lubricity.
Due to lattice mismatch, the interfacial energy landscape is exceptionally flat, and friction arises predominantly from weak coupling between collective vibrational modes across the incommensurate interface.
Although long-wavelength Moir\'e patterns may develop~\cite{Gao2022FC}, the associated lateral energy corrugation ($\sim$1 meV per atom) remains far below $k_{\rm B}T$ at 300 K, precluding static pinning and suppressing local mechanical instabilities.

\begin{figure}[ht]
\centering    
\includegraphics[width=0.8\linewidth]{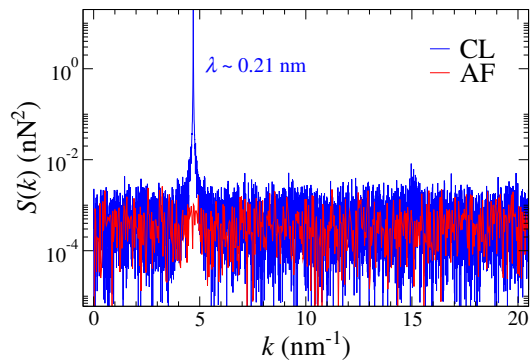}
\caption{
Spatial power spectral density $S(k)$ of the lateral force for the area-filling (AF) and contact-line (CL) geometries.
In both cases, the Au(111) slab slides along the graphene armchair direction at a velocity of 10 m/s under a normal load of 0.1 MPa.
}
\label{fig:psd}
\end{figure}

In contrast, finite contacts with explicit terminations (contact-line, CL geometry) break lateral translational symmetry and introduce enhanced compliance near the interface edge.
Dissipation therefore becomes spatially heterogeneous and is locally amplified in the vicinity of the contact line.
This distinction is directly reflected in the spatial power spectral density (PSD) of the lateral force, shown in Fig.~\ref{fig:psd}.
The PSD is computed from the zero-mean force signal $F(x)$ as
\begin{equation}
S(k)=\frac{1}{N}\sum_{n=0}^{N-1} |F(x_n)e^{ikx_n}|^2,
\label{eqn:psd}
\end{equation}
where $x_n=n\Delta x$ and $k$ denotes the spatial frequency.

For the AF interface, the PSD exhibits no pronounced spectral peak, consistent with the smooth sliding characteristic of structural lubricity and the absence of coherent atomic-scale force modulation. 
In contrast, the CL geometry displays a distinct peak at $k\approx4.7$ nm$^{-1}$, corresponding to a real-space wavelength $\lambda=1/k\approx0.21$ nm.
This value matches the graphene armchair periodicity ($\sqrt{3}a/2,~a=0.246$ nm), indicating lattice-resolved stick-slip motion localized at the contact termination.

\begin{figure}
\centering    
\includegraphics[width=0.8\linewidth]{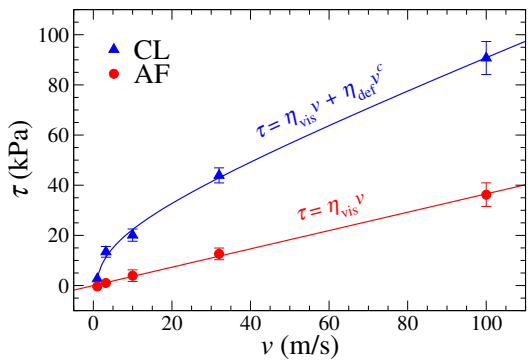}
\caption{
Interfacial shear stress $\tau$ as a function of sliding velocity $v$ for the area-filling (AF) and contact-line (CL) geometries at a normal load of 0.1 MPa.
The AF interface exhibits a linear viscous response, $\tau=\eta_{\rm vis}v$ with $\eta_{\rm vis}=0.365$ kPa$\cdotp$s/m.
In contrast, the CL geometry follows the composite form
$\tau(v)=\eta_{\rm vis}v+\eta_{\rm def}v^c$,
with $\eta_{\rm vis}=0.090$ kPa$\cdotp$s/m, $\eta_{\rm def}=5.03$ kPa$\cdotp$(s/m)$^c$, and $c=0.61$.
}
\label{fig:fx_v}
\end{figure}

The velocity dependence of the interfacial shear stress $\tau$ further highlights the role of geometry, as shown in Fig.~\ref{fig:fx_v}.
For the AF interface, the shear stress is orders of magnitude smaller than in the finite-contact case and scales linearly with sliding velocity, 
\begin{equation}
\tau=\eta_{\rm vis}v,
\end{equation}
defining an effective viscous coefficient $\eta_{\rm vis}$.
This linear scaling contrasts sharply with barrier-controlled sliding, in which friction is only weakly velocity dependent and approaches Coulomb-like behavior at low speeds~\cite{Mueser2011PRB}.
The observed Stokesian response therefore confirms that dissipation in the AF geometry is not governed by stick-slip activation over static energy barriers.
This behavior is insensitive to thermostat choice and persists across Langevin damping times from 0.1 to 10 ps as well as under Nos\'e–Hoover dynamics.

In the CL geometry, the shear stress is substantially larger and exhibits a weakened, sublinear dependence on sliding velocity.
This behavior reflects the coexistence of two dissipation channels:
viscous phononic damping~\cite{Riva2025PRB}, present in both geometries, and localized stick-slip processes concentrated near the contact termination.
Over the explored velocity range, the rate dependence is well captured by
\begin{equation}
\tau(v)=\eta_{\rm vis}v+\eta_{\rm def}v^c,
\label{eqn:vis_act_eqn}
\end{equation}
where the linear term represents phonon-mediated damping and the sublinear contribution captures excess edge-induced dissipation. 
The exponent $c<1$ signifies weak rate sensitivity, consistent with thermally activated barrier-crossing processes, which in simplified models often produce logarithmic velocity dependence but may be approximated by an effective power law over finite ranges.

The smaller fitted value of $\eta_{\rm vis}$ in the CL geometry cannot be attributed solely to the reduced contact area. 
Although geometric scaling would modestly lower the viscous stress when normalized by a smaller projected area, the substantial decrease of $\eta_{\rm vis}$ exceeds what is expected from area effects along. 
This indicates that the finite contact does not simply weaken phonon-mediated coupling in the interior; rather, dissipation becomes spatially heterogeneous:
bulk-like viscous damping persists within the contact interior, while additional edge-localized, rate-dependent processes contribute to the total shear stress.
The fitted viscous coefficient therefore represents an effective parameter arising from the superposition of distinct dissipation channels, rather than an intrinsic material constant.

\begin{figure}[ht]
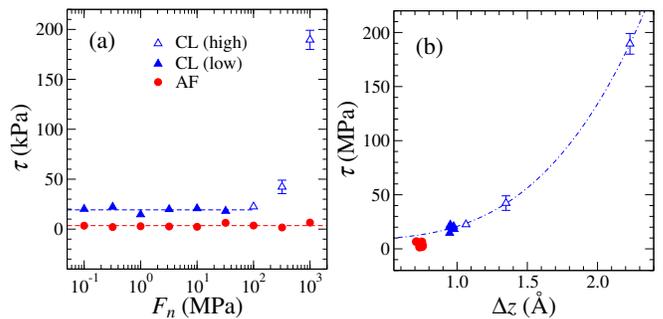

\centering    
\includegraphics[width=0.49\linewidth]{Figure/load_tau.eps}
\includegraphics[width=0.49\linewidth]{Figure/dz_stress.eps}
\caption{
Interfacial shear stress $\tau$ as a function of (a) normal load $F_n$ and (b) out-of-plane deformation amplitude $\Delta z$ of the top graphene layer for the area-filling (AF) and contact-line (CL) geometries.
Solid and open triangle symbols denote the low-load ($F_n<$100 MPa) and high-load ($F_n\geq$100 MPa) regimes, respectively.
%
%
The dashed curve in (b) represents a fit to the CL data using Eq.~\ref{eqn:stress_dz}, yielding $\beta=3.4$.
Error bars smaller than the symbol size are omitted.
}
\label{fig:dz_stress}
\end{figure}

A striking result emerges when the shear stress is examined as a function of normal load.
For the AF interface, the interfacial shear stress remains essentially constant over the entire investigated range from 0.1 MPa to 1 GPa (solid red circles in Fig.~\ref{fig:dz_stress}a). 
Such load invariance stands in sharp contrast to classical frictional contacts, where increasing load enhances friction through contact-area growth or load-induced amplification of lateral energy barriers. 
In the present incommensurate and laterally infinite geometry, the real contact area is fixed and static pinning barriers are absent.
Normal pressure therefore neither reshapes the interfacial energy landscape nor activates new mechanisms of dissipation.

%
%
%
%

To clarify the origin of this invariance, we examine the steady-state energy balance during sliding.
At velocity $v$, the shear stress may be written as
\begin{equation}
\tau=\frac{P_{\rm diss}}{Av},
\end{equation}
where $P_{\rm diss}$ is the rate of energy dissipation and $A$ the contact area.
For incommensurate AF interfaces, dissipation is governed by phonon-mediated interfacial damping that scales as
$P_{\rm diss}=\Gamma_{\rm int}v^2$.
This yields
\begin{equation}
\tau=\frac{\Gamma_{\rm int}}{A}v\equiv\eta_{\rm vis}v,
\end{equation}
identifying $\eta_{\rm vis}=\Gamma_{\rm int}/A$ as an effective viscous coefficient.
The observed load independence thus implies that the interfacial damping constant $\Gamma_{\rm int}$ remains essentially invariant with normal pressure.
Establishing $\Gamma_{\rm int}$ as a load-insensitive parameter demonstrates that friction in this regime is governed by intrinsic dynamical coupling across the interface, rather than by load-modulated geometric or barrier effects.

In the CL geometry, the shear stress is approximately five times larger than in the AF case due to additional edge-localized dissipation.
Remarkably, however, it remains nearly load independent up to $\sim$100 MPa (solid blue triangles in Fig.~\ref{fig:dz_stress}a). 
The presence of a contact termination therefore increases the magnitude of friction without immediately introducing load sensitivity.
Throughout this regime, the interface remains atomically conformal, with no evidence of plastic deformation or structural rearrangement.

A pronounced deviation from load-independent behavior emerges only at higher normal stresses ($\gtrsim$300 MPa), where the shear stress rises rapidly.
This increase correlates with enhanced out-of-plane bending of the graphite substrate localized near the contact termination.
The deformation amplitude is quantified by the vertical height variation of the top graphene layer,
\begin{equation}
\Delta z=z_{\rm max}-z_{\rm min}.
\end{equation}
As shown in Fig.~\ref{fig:dz_stress}b, $\Delta z$ remains nearly constant ($\sim$1.0 \AA) throughout the load-independent regime but increases sharply beyond a critical threshold $\Delta z_{\rm c}$, coincident with the onset of enhanced friction.

Unlike classical load-induced friction increases that stem from amplification of lateral energy barriers at individual asperities~\cite{Socoliuc2004PRL}, the present transition does not involve the emergence of static pinning. 
The interface remains incommensurate across the entire load range.
Instead, the breakdown of load independence originates from elastic bending concentrated at the contact termination.
For small deflections, the bending energy of a thin sheet scales approximately as
\begin{equation}
E_{\rm bend}\propto (\Delta z)^2,
\end{equation}
reflecting its quadratic dependence on curvature. 
While comparable vertical fluctuations may also occur in the AF geometry (e.g., solid red circles in Fig.~\ref{fig:dz_stress}b), they do not produce a similar stress increase, indicating that localized compliance at the contact line, rather than bending amplitude alone, is responsible for the additional dissipation.

To characterize the onset of enhanced dissipation, we define the excess deformation
\begin{equation}
\delta z=\Delta z-\Delta z_{\rm c}.
\end{equation}
Once $\delta z>0$, elastic bending alters local mechanical compliance and strengthens vertical-lateral coupling, activating additional dissipation channels.
The shear stress can therefore be expressed phenomenologically as
\begin{equation}
\tau(\Delta z) = \tau_0
+ K_z\delta z^\beta\Theta(\delta z),
\label{eqn:stress_dz}
\end{equation}
where $\tau_0$ denotes the load-independent baseline stress, $K_z$ characterizes deformation-induced dissipation, $\beta>1$ captures the superlinear activation of bending-mediated coupling, and $\Theta(x)$ is the Heaviside step function.
The superlinear exponent reflects the combined effects of quadratic bending energy storage and nonlinear interfacial coupling.
This formulation captures both the persistence of load-independent friction below $\Delta z_{\rm c}$ and the rapid activation of additional dissipation once localized compliance exceeds the critical threshold.
Notably, this transition occurs without plasticity or structural transformation, demonstrating that elastic edge compliance along governs the breakdown of load-independent friction in finite contacts.

In summary, we show that genuine load-independent friction can emerge in structurally lubric interfaces, but only within well-defined geometric and elastic constraints. 
For laterally infinite, area-filling Au(111)/graphite contacts, sliding follows a linear viscous velocity dependence and remains insensitive to normal load over three orders of magnitude in pressure, consistent with phonon-mediated interfacial damping as the dominant dissipation mechanism. 
When the contact includes an explicit termination line, the friction level increases markedly and the velocity dependence becomes sublinear, reflecting additional edge-localized dissipation.
Even in this finite geometry, however, friction remains load independent up to a critical stress.
Only when out-of-plane elastic bending at the contact termination exceeds a threshold does the shear stress rise rapidly.

These results show that load independence is not guaranteed by incommensurability alone, but instead represents a geometry- and compliance-limited dynamical state.
Its breakdown does not require plastic deformation, disorder, or the onset of static pinning.
Rather, it originates from localized elastic bending that enhances vertical-lateral coupling and opens additional dissipation pathways.
Structural lubricity therefore persists only when dissipation remains bulk-like and edge-induced compliance is suppressed.
In this way, the transition from ideal infinite interfaces to realistic finite contacts can be understood as a controlled shift from uniform interfacial damping to deformation-mediated energy loss.

\begin{acknowledgments}
HG thanks Martin H. M\"user for useful discussions.
This work was supported by the German Research Foundation (DFG) under Grant No. GA 3059/4-1.
\end{acknowledgments}

\bibliography{achemso-demo}

\end{document}